\documentclass[fleqn,usenatbib]{mnras}

\usepackage{newtxtext,newtxmath}

\usepackage[T1]{fontenc}
\usepackage{ae,aecompl}

\usepackage{graphicx}	
\usepackage{amsmath}	
\usepackage{bm}
\usepackage[colorinlistoftodos]{todonotes}
\usepackage{ dsfont }

\DeclareMathOperator{\Lie}{\mathcal{L}}

\DeclareMathOperator{\x}{\text{x}}
\DeclareMathOperator{\n}{\text{n}}
\DeclareMathOperator{\p}{\text{p}}
\DeclareMathOperator{\vort}{\text{L}}

\def\eul{{\mathfrak{n}}}

\title[Vortex pinning in relativistic neutron stars]{Vortex pinning in the superfluid core of relativistic neutron stars }

\author[A. Sourie and N.~Chamel]{
Aur\'elien Sourie,$^{1,2}$\thanks{E-mail: aurelien.sourie@gmail.com}
and Nicolas Chamel$^{1}$\thanks{E-mail: nchamel@ulb.ac.be}
\\
$^{1}$Institut d'Astronomie et d'Astrophysique,  Universit\'e Libre de Bruxelles, CP-226, B-1050 Brussels,
Belgium\\
$^{2}$Laboratoire Univers et Th\'eories, Observatoire de Paris, Universit\'e PSL, CNRS, Universit\'e de Paris, F-92190 Meudon, France
}

\date{Accepted XXX. Received YYY; in original form ZZZ}

\pubyear{2021}

\begin{document}
\label{firstpage}
\pagerange{\pageref{firstpage}--\pageref{lastpage}}
\maketitle

\begin{abstract}
Our recent Newtonian treatment~\citep{sourie20force,sourie20glitch} of the smooth-averaged  mutual-friction force acting on the neutron superfluid and locally induced by the pinning of quantized neutron vortices to proton fluxoids in the outer core of superfluid neutron stars is here adapted to the general-relativistic framework. We show how the local nonrelativistic motion of individual vortices can be matched to the global dynamics of the star using the fully 4D covariant Newtonian formalism of \cite{carterchamel2004}. 
We derive all the necessary dynamical equations for carrying out realistic simulations of superfluid rotating neutron stars in full general relativity, as required for the  interpretation of pulsar frequency glitches. The role of vortex pinning on the global dynamics appears to be nontrivial.
\end{abstract}

\begin{keywords}
stars: interiors, stars: neutron
\end{keywords}

\section{Introduction}

Pulsar frequency glitches~\citep{manchester2018} are peculiar astrophysical phenomena that are thought to reveal the existence of superfluidity~\citep{chamel2017} in the interior of neutron stars (NSs), the cold and dense remnants of gravitational (core-collapse) supernova explosions. The sudden spins up and the subsequent long relaxations, as observed in the emblematic  Vela pulsar, were originally explained by the unpinning and creep of neutron quantized vortices in the neutron-star crust~\citep{anderson1975,alpar1984vortex,alpar1984rapid}. However, the details of the vortex dynamics and the stellar regions involved during glitches still remain uncertain (see, e.g., \cite{haskell2015models,graber2017,haskell2018} for recent reviews). Indeed, it has been found that the presence of inhomogeneities in the crust tends to suppress superfluidity~\citep{chamel2017JLTP,watanabe2017,sauls2020}, which may thus play a less important role than initially thought~\citep{andersson2012pulsar,chamel2013crustal,delsate2016}. On the other hand, angular momentum can also be stored in the superfluid core and different alternative astrophysical scenarios have been proposed~\citep{sedrakian1999,jahan-miri2002,peralta2006,pizzochero2011,ho2015,pizzochero2020}.

In particular, neutron vortices may pin to proton fluxoids in the core of NSs \citep{muslimov1985vortex,sauls1989superfluidity} (see also \cite{alpar2017} for a recent review), considering protons form a type$-$II superconductor, as first argued by \cite{baym1969superfluidity}. Because a toroidal magnetic field is expected to be  present in the outer core of a NS, in the region beneath the crust (see, e.g. \cite{ankan2020} and references therein), vortex pinning is unavoidable. Therefore that region of the core also contributes to  glitches and their relaxation~\citep{gugercinoglu2014vortex,gugercinoglu2017,gugercinoglu2020}. Although vortices may cut through fluxoids depending on their velocity and on the pinning strength, \cite{ruderman1998neutron} estimated that this does not occur in Vela-like pulsars. The pinning of vortices to fluxoids may also drive crustal plate tectonics and play a key role in the evolution of the magnetic field~\citep{srinivasan1990novel,ruderman1998neutron}.  Alternatively, \cite{sedrakian1995superfluid} argued that fluxoids could actually be naturally nucleated in the vicinity of each vortex, thus forming ``vortex clusters''. 
In either case, we have 
shown~\citep{sourie20glitch} that the rigid motion of vortices and fluxoids 
could explain specific timing features that have been recently observed in the Crab and Vela pulsars~\citep{shaw2018largest,palfreyman18alteration,asthon19rotational}. 

Our analysis was carried out in the framework of  Newtonian theory. Although the motions of individual vortices are locally nonrelativistic, their  typical velocities in the core are of order of 1~cm~s$^{-1}$~\citep{gugercinoglu2016microscopic}, the smooth-averaged hydrodynamics of the  superfluid at the global scale of the star is prone to general-relativistic effects, especially in the most massive NSs (see, e.g.,~\cite{sourie2017global, gavassino20} in the context of glitches). In particular, frame dragging in rotating NSs induces additional couplings between the superfluid and the rest of star, as first discussed by \cite{carter1975}. We have shown that the ensuing coupling coefficients may be of comparable magnitude (although of opposite sign) as those due to the mutual entrainment induced by nuclear interactions~\citep{sourie2017global}. 

In this paper, we adapt our recent model of superfluid NSs~\citep{sourie20force,sourie20glitch} to the general-relativistic framework. To this end, we extend the analysis of~\cite{langlois1998differential} to allow for the pinning of vortices to fluxoids or the formation of vortex clusters in the outer core of NSs, based on the general formalism of dissipative superfluid mixtures developed by \cite{carter2005covIII}  in the Newtonian context. Making use of the fully 4D covariance of this formalism, we also show how the local nonrelativistic vortex dynamics can be matched to the global hydrodynamic description of the star. Our model of superfluid NS is presented in Section~\ref{sec:two-fluid}. Applications to quasi-stationary rotating NSs, including the calculation of the mutual-friction force, are discussed in Section~\ref{sec:mutual-friction}. Unless stated otherwise, we shall set the speed of light $c=1$.

\section{Covariant two-constituent superfluid hydrodynamics}
\label{sec:two-fluid}

\subsection{Carter-Langlois-Sedrakian two-fluid model}

In this section, we shall briefly review the mean features of the two-fluid model of \cite{langlois1998differential} (for a general description of superfluid NS, see e.g. ~\cite{glampedakis2011magnetohydro, GD2016relativistic} and references therein). 

Since electrically-charged particles inside NSs are strongly coupled and essentially corotate with the crust and the magnetosphere, 
the outer core of a cold mature NS can be reasonably well described in terms of just two dynamically distinct fluids, namely (i) an inviscid neutron superfluid with 4-current $n_{\n}^{\mu}= n_{\n}\, u_{\n}^{\mu}$ and (ii) a fluid made of protons and electrons with 4-current $n_{\p}^{\mu}=  n_{\p}\, u_{\p}^{\mu}$, $u_{\n}^{\mu}$ and $u_{\p}^{\mu}$ denoting the corresponding 4-velocities. This latter component will be referred to as the `normal' fluid throughout the paper. 
In what follows, not only will the total baryon 4-current 
\begin{equation}
n_\text{b}^{\mu} = n_{\n}^{\mu} + n_{\p}^{\mu} 
\end{equation}
be conserved, i.e., 
\begin{equation}
\nabla_{\mu} n_{\text{b}}^{\mu} = 0\, ,    
\end{equation}
but we will also neglect any kind of transfusive processes whereby one constituent is converted into the other.
Each 4-current is therefore assumed to be separately conserved\footnote{Although such processes could be readily incorporated to the present model~\citep{langlois1998differential}, their impact on the superfluid dynamics of NSs is expected to be very weak (see, e.g.,~\cite{sourie2017global}).},
\begin{equation}
\nabla_{\mu} n_{\n}^{\mu} = 0 \ \ \text{and} \  \   \nabla_{\mu} n_{\p}^{\mu} = 0\, .  
\label{eq:cons_num}
\end{equation}
Following \cite{langlois1998differential}, 
the entropy current $s^{\mu}$ is not treated as an independent fluid,  but is assumed to be expressible as 
\begin{equation}
s^{\mu} = s \, u_{\p}^{\mu} \, , 
\end{equation}
where $s$ is the entropy density in the rest-frame of the charged particles. 

The local thermodynamic state of the system under consideration can be described by a Lagrangian density $\Lambda$, commonly referred to as the master function, which depends on both particle 4-current vectors $n_{\n}^{\mu}$ and $n_{\p}^{\mu}$ and on the entropy density~$s$,
\begin{equation}
\Lambda\left(n_{\n}^{\mu}, n_{\p}^{\mu}, s \right)\, .
\end{equation}
Variations of this quantity (keeping fixed the space-time metric\footnote{Variations of the total Lagrangian density (including the Einstein-Hilbert contribution) with respect to the space-time metric yield Einstein's equations.}) lead to 
\begin{equation}
\delta \Lambda = - \Theta\, \delta s +  p_{\mu}^{\n} \,\delta n_{\n}^{\mu}+  p_{\mu}^{\p}\, \delta n_{\p}^{\mu}\, , 
\end{equation}
where $\Theta$ is interpretable as the thermodynamic temperature of the system as measured in the rest frame of the normal fluid, and $p_{\mu}^{\n}$ (resp. $p_{\mu}^{\p}$) denotes the canonical 4-momentum per baryon of the neutron superfluid (resp. the normal fluid). Let us remark that, due to non-dissipative entrainment effects arising from the nuclear interactions between neutrons and protons, the 4-momentum of a given fluid is not simply aligned with its corresponding 4-velocity but also depends on the 4-velocity of the second fluid (see, e.g., \cite{ gusakov2014,sourie2016numerical,leinson2018, chamel2019} for recent calculations of the coupling coefficients). 

Carter and collaborators have developed an elegant action principle to derive the fluid equations of motion from the Lagrangian density~$\Lambda$ by considering variations of the fluid particle trajectories (see, e.g., \cite{carter1989covariant, langlois1998differential, carter1998relativistic} for details; see also \cite{andersson2020} for a review). 
Applied to the two-fluid model under consideration here, this procedure yields the following expression for the energy-momentum tensor of the system: 
\begin{equation}
\label{eq:Tmunu}
T^{\mu}_{\nu} = \Psi \delta^{\mu}_{\nu} + s^\mu \Theta_\nu  + n_{\p}^{\mu}p^{\p}_{\nu} + n_{\n}^{\mu}p^{\n}_{\nu}\, , 
\end{equation}
where $\Psi$ denotes the generalized pressure of the fluids, and $\Theta_\nu$ is the thermal 4-momentum per ``entropon'' as referred to by Carter, i.e., the 4-momentum per one unit of entropy dynamically conjugate to the entropy current, see e.g. equation (18) of~\cite{langlois1998differential}. Note that the temperature can be alternatively interpreted as the chemical potential of entropons in the thermal rest frame \citep{carter1989covariant}
\begin{equation}\label{eq:temperature}
    \Theta=-u_{\p}^\mu \Theta_\mu \, .
\end{equation}
This approach also leads to the following force laws [see equations (20) and (21) of \cite{langlois1998differential}]
\begin{equation}
\label{force_volp}
 f^{\p}_\mu = n_{\p}^{\nu} \varpi^{\p}_{\nu \mu} 
\end{equation}
and 
\begin{equation}
\label{force_voln}
f^{\n}_\mu = n_{\n}^{\nu} \varpi^{\n}_{\nu \mu} \, ,
\end{equation}
where (square brackets denoting antisymmetrization)
\begin{equation}
\label{vorticity_2-form}
\varpi^{\x}_{\nu \mu}  = 2\,  \nabla_{[\nu}p^{\x}_{\mu]}= \nabla_{\nu}p^{\x}_{\mu}- \nabla_{\mu}p^{\x}_{\nu}
\end{equation}
stands for the vorticity 2-form 
averaged over scales larger than the intervortex separation (see below). The 4-covectors $f^{\p}_\mu$ and $f^{\n}_\mu$ involved in equations~\eqref{force_volp} and~\eqref{force_voln} are to be interpreted as the mean force densities acting on the normal fluid and the neutron superfluid, respectively. 

At sufficiently small scale (but large enough for the hydrodynamic description to remain valid), the 4-momentum $p^{\n}_{\mu}$ of the neutron superfluid is given by the gradient of the phase $\phi$ of the quantum condensate (see, e.g., \cite{carter1998relativistic}), 
\begin{equation}
\label{eq:gradient_phase}
    p^{\n}_{\mu} = \dfrac{m_{\n}\kappa_{\n}}{2\pi}\nabla_{\mu} \phi\, , 
\end{equation}
where $\kappa_{\n} = h/(2 m_{\n})$, $h$ being the Planck constant and $m_{\n}$ the neutron rest mass. This relation implies that the superflow is irrotational, as characterized by the vanishing of the corresponding vorticity 2-form~\eqref{vorticity_2-form}. Nevertheless, it is well-known from laboratory experiments (see, e.g., \cite{yarmchuk1979observation}) that the condition~\eqref{eq:gradient_phase} can be locally violated through the formation of quantized vortices, each carrying a quantum of circulation~$\kappa_{\n}$. 

The existence of such vortex filaments leads to the non-vanishing of the macroscopically-averaged (i.e., averaged on scales much larger than the mean intervortex separation) vorticity 2-form $\varpi^{\n}_{\mu \nu}$. However, the underlying presence of quantized vortices implies that $\varpi^{\n}_{\mu \nu}$ must be of rank 2 instead of 4 (see, e.g., \cite{carter1989covariant,langlois1998differential,carter2001relativistic,andersson2016quantised,gavassino20} for further discussions).

\subsection{Interactions between vortices and the surrounding fluids}
\label{sec:vortex-interaction}

Having recalled the main features of the Carter-Langlois-Sedrakian two-fluid model, as well as the conditions imposed by  superfluidity, the second law of thermodynamics can now be invoked to constrain the expression for the force density $f^{\n}_\mu$~\eqref{force_voln} exerted on the neutron superfluid  by the vortex lines. A seminal work towards this direction was initiated by~\cite{langlois1998differential} (see also \cite{andersson2016quantised, gavassino20} for similar studies). 
We shall extend the treatment of~\cite{langlois1998differential} by following the more general approach of dissipation in superfluid mixtures developed by  \cite{carter2005covIII} in a fully 4D covariant Newtonian framework. 

Let us first introduce the thermal 4-force density defined by
\begin{equation}
    \label{force_ent}
    f^{\emptyset}_\mu =  2\, s^{\nu}\, \nabla_{\left[\nu\right.} \Theta_{\left.\mu\right]} + \Theta_{\mu} \, \nabla_{\nu} s^{\nu}\, .
\end{equation}
In the strictly conservative case, as characterized by the absence of any force other than those already included in the Lagrangian description, the separate force densities $f_{\p \mu}$, $f_{\n \mu}$, and $f^{\emptyset}_\mu$ must all vanish. 
However, in the more general (dissipative) context under interest here, the force densities will be 
subject to the following relation~\citep{carter2005covIII}
\begin{equation}
\label{fext}
f^{\n}_\mu  + f^{\p}_\mu +f^{\emptyset}_\mu = f_{\text{ext}\,\mu}\, , 
\end{equation}
where $f_{\text{ext}\,\mu}= \nabla_{\nu} T^{\nu}_{\mu}$ denotes the total external force density acting on the system, which arises from the loss of rotational energy through the emission of electromagnetic radiation and from internal heating. It can thus be decomposed into two forces acting separately on the charged and on the entropy components,
\begin{equation}
\label{fext2}
    f_{\text{ext}\,\mu}=f^{\p}_{\text{ext}\,\mu}+f^{\emptyset}_{\text{ext}\,\mu} \, ,
\end{equation}
considering that the neutron superfluid is not directly subject to any external force.

Without any loss of generality, the neutron force density $f^{\n}_\mu $ can be decomposed as
\begin{equation}
    f^{\n}_\mu  = f^{\n}_{\text{d}\,\mu}  + f^{\n}_{\text{c}\,\mu}\, , 
\end{equation}
where $f^{\n}_{\text{d}\,\mu}$ (resp. $f^{\n}_{\text{c}\,\mu}$) is a dissipative (resp. conservative) force term, usually referred to as `drag' (resp. `transverse') force in the literature. 
Using this decomposition, equation~\eqref{force_voln} now reads
\begin{equation}
\label{force_voln_bis}
 n_{\n}^{\nu} \varpi^{\n}_{\nu \mu} =  f^{\n}_{\text{d}\,\mu}  + f^{\n}_{\text{c}\,\mu}\, .
\end{equation}
The conservative force term $f^{\n}_{\text{c}\,\mu}$ was ignored by \cite{langlois1998differential}, as can be seen from their equation (41). 
The simplest prescription to ensure that Eq.~\eqref{force_voln_bis} is compatible with the existence of two null eigenvectors, say $w_1^\mu$ and $w_2^\mu$, for the neutron vorticity 2-form $\varpi^{\n}_{\mu\nu}$ (as implied by the presence of quantized vortices), is to require that each force density covector is orthogonal to both $w_1^\mu$ and $w_2^\mu$:
\begin{equation}
\label{eq.force.worldsheet}
w_1^\mu f^{\n}_{\text{d}\,\mu}=0=w_2^\mu  f^{\n}_{\text{d}\,\mu} \ \ \text{and} \ \ w_1^\mu f^{\n}_{\text{c}\,\mu}=0=w_2^\mu  f^{\n}_{\text{c}\,\mu} \, ,
\end{equation}
where by definition
\begin{equation}\label{eq.null.eigenvectors}
w_1^\mu \varpi^{\n}_{\mu\nu}=0=w_2^\mu  \varpi^{\n}_{\mu\nu} \, .
\end{equation}
In other words, $f^{\n}_{\text{d}\,\mu}$ and $f^{\n}_{\text{c}\,\mu}$ must be orthogonal to the two-dimensional vortex worldsheet swept by the vectors $w_1^\mu$ and $w_2^\mu$. 
Introducing the corresponding (orthogonal) projector $\perp^\nu_\mu$, we must have 
\begin{equation}
\label{eq.force.worldsheet2}
\perp_\mu^\nu f^{\n}_{\text{d}\,\nu}=f^{\n}_{\text{d}\,\mu} \ \ \text{and} \ \ \perp_\mu^\nu f^{\n}_{\text{c}\,\nu}=f^{\n}_{\text{c}\,\mu} \, .
\end{equation}
By definition, 
\begin{equation}
\label{eq.force.worldsheet3}
\perp_\mu^\nu \varpi^{\n}_{\nu \rho} = \varpi^{\n}_{\mu\rho} \, .
\end{equation}
Contracting Eq.~\eqref{eq.force.worldsheet3} again by the projector leads to 
\begin{equation}
\perp_\nu^\mu \perp^\nu_\rho\,=\,\perp^\mu_\rho \, .
\end{equation}
In view of Eq.~\eqref{eq.null.eigenvectors}, we also have 
\begin{equation}
\perp_\nu^\mu w_1^\nu = 0 =\, \perp_\nu^\mu w_2^\nu \, .
\end{equation}
The explicit form of the projector will be given in the next subsection, see Eq.~\eqref{perp}. 
Since the vorticity is carried by vortex lines, one of the null eigenvectors $w_1^\mu$ or $w_2^\mu$ must be 
timelike, see Appendix~\ref{app:vel_vort} for further discussions. The orthogonality conditions~\eqref{eq.force.worldsheet} thus imply that $f^{\n}_{\text{d}\,\mu}$ and $f^{\n}_{\text{c}\,\mu}$ are spacelike.

The actual forms of $f^{\n}_{\text{d}\,\mu}$ and $f^{\n}_{\text{c}\,\mu}$ can be deduced from the second law of thermodynamics, which can be locally expressed as 
(see, e.g., equation (23) of \cite{carter2005covIII})
\begin{equation}\label{second_law}
\Theta  \nabla_{\mu}  s^{\mu} + \mathcal{Q}
\geq 0 \, ,
\end{equation}
where $\mathcal{Q}=u_{\p}^\mu f^{\emptyset}_{\text{ext}\,\mu}$ denotes the heat loss rate per unit volume in the thermal rest frame. In the following, we shall adopt the weak closure condition according to which the external force density acting on the normal fluid does not vanish but is restricted by the following relation  \citep{carter2005covIII} 
\begin{equation}
\label{weak-closure}
u_{\p}^{\mu}f^{\p}_{\text{ext}\,\mu}=0\, . 
\end{equation}
In view of Eqs.~\eqref{force_volp} and \eqref{force_voln}, we find the similar relations 
\begin{equation}
\label{weak-closure2}
 u_{\n}^{\mu}f^{\n}_{\mu}=0 \ \ \text{and} \ \ u_{\p}^{\mu}f^{\p}_{\mu}=0\, . 
\end{equation}
Contracting Eq.~\eqref{fext2} with $u_{\p}^\mu$ using Eqs.~\eqref{eq:temperature}, \eqref{force_ent}, \eqref{fext}, \eqref{weak-closure}, and \eqref{weak-closure2} yields
\begin{equation}\label{entropy_balance}
\Theta  \nabla_{\mu}  s^{\mu} + \mathcal{Q} = u_{\p}^\mu f^{\n}_{\mu}\, .
\end{equation}
Since the force density $f^{\n}_{\text{c}\,\mu}$ does not lead to any dissipation, the following relation should hold:
\begin{equation}
\label{non_diss}
u_{\p}^{\mu}  f^{\n}_{\text{c}\,\mu} = 0\, , 
\end{equation}
so that Eq.~\eqref{entropy_balance} reduces to 
\begin{equation}
\label{entropy_balance2}
\Theta  \nabla_{\mu}  s^{\mu} + \mathcal{Q} = u_{\p}^\mu f^{\n}_{\text{d}\,\mu} \, .
\end{equation}
An obvious way to make sure that $f^{\n}_{\text{c}\,\mu}$ satisfies both Eqs.~\eqref{eq.force.worldsheet} and~\eqref{non_diss} is to write 
\begin{equation}
     f^{\n}_{\text{c}\,\mu} = \mathcal{A} \,   \varpi^{\n}_{\mu\nu}   n_{\p}^{\nu}\, , 
    \label{force_nondiss}
\end{equation}
with some unknown coefficient $\mathcal{A}$. On the other hand, the second law of thermodynamics~\eqref{second_law} requires that $u_{\p}^{\mu}   f^{\n}_{\text{d}\,\mu}$ be positive definite, or equivalently $u_{\p \perp}^\mu f^{\n}_{\text{d}\,\mu}>0$ with $u_{\p \perp}^\mu\equiv \, \perp^\mu_\nu u_{\p}^\nu$ in view of Eq.~\eqref{eq.force.worldsheet2}. This condition can be fulfilled by expressing the dissipative force as 
\begin{equation}
f^{\n}_{\text{d}\,\mu} =  \mathcal{C}_r \, u_{\p\perp\, \mu}\,  ,
 \label{drag_force}
\end{equation}
(recalling that $f^{\n}_{\text{d}\,\mu}$ hence also $u_{\p\perp}^\mu$ are spacelike), 
where $\mathcal{C}_r$ is a positive coefficient. 
For later convenience, we introduce the (positive) resistivity coefficient $\mathcal{R}= \mathcal{C}_r  / \mathcal{N}_{\n}$, where $\mathcal{N}_{\n}$ is the average surface density of vortex lines. The force balance equation~\eqref{force_voln_bis} can finally be expressed as 
\begin{equation}
\label{FORCE_BALANCE}
 n_{\n}^{\nu} \varpi^{\n}_{\nu \mu} =  \mathcal{N}_{\n}\,   \mathcal{R}\, u_{\p \perp\, \mu} + \mathcal{A}\,   \varpi^{\n}_{\mu\nu}   n_{\p}^{\nu}\, , 
\end{equation}
where, up to this point, $\mathcal{R}$ and $\mathcal{A}$ are left unspecified. 
Equation~(\ref{FORCE_BALANCE}) represents the mutual-friction force, i.e., the effective force density exerted by the normal fluid on the neutron superfluid due to the average forces acting on individual vortices. This expression is very general since we have made no assumption on the spatial arrangement of vortex lines. \cite{langlois1998differential} implicitly assumed that $\mathcal{A}=0$. However, as we shall show in the next section, this coefficient is nonzero whenever proton fluxoids are pinned to vortices or vortex clusters of the kind proposed by \cite{sedrakian1995superfluid} are formed.

\subsection{Matching between the local and global dynamics}

The coefficients $\mathcal{R} $ and $\mathcal{A}$ appearing in equation~\eqref{FORCE_BALANCE} should be provided by an analysis of the local perturbations of the fluid flows induced by the motion of  individual vortices. At such small scales, the space-time curvature is negligible (see, e.g., the  discussion in Section 3.4 of \cite{glendenning1997}) and the fluid dynamics is essentially nonrelativistic. We have recently derived the Newtonian expression for the force per unit length exerted on a single vortex line in a mixture of superfluid neutrons, superconducting protons and degenerate electrons, as can be typically found in the outer core of NSs~\citep{sourie20force}. Allowing for the possible presence of vortex pinning onto proton fluxoids or vortex clusters of the kind proposed by \cite{sedrakian1995superfluid}, we considered the force acting on a vortex line to which a given number $N_{\p}$ of fluxoids are anchored\footnote{The $N_{\p}$ fluxoids are not superimposed but lie in the vicinity of the vortex, 
within distances much smaller than the intervortex spacing.}. 
This  number could be huge, of order $\sim 10^{13}$ \citep{sourie20glitch}. 
The resulting expression for the force acting in a vortex involves an unspecified positive dimensionless coefficient $\xi$, the so-called drag-to-lift ratio, which measures the amplitude of the microscopic drag force acting on each vortex line (which is thought to arise from the scattering of electrons off the magnetic field carried by the quantized lines). Although the drag-to-lift ratio associated with a vortex line pinned to $N_{\p}$ fluxoids is essentially unknown (see~\cite{sourie20glitch}), this parameter is expected to depend on $N_{\p}$. The smooth-averaged force per unit volume exerted on the neutron superfluid at the macroscopic scale, as induced by the drag forces acting on individual vortices, was derived in~\cite{sourie20glitch} within the Newtonian framework considering straight and infinitely rigid vortices, assumptions that remain applicable in general relativity~\citep{gavassino20}. One can now use these results to determine the expressions for $\mathcal{R}$ and $\mathcal{A}$.

To find the Newtonian limit of the force balance equation~\eqref{FORCE_BALANCE}, we rely on the fully 4D-covariant Newtonian formulation developed by \cite{carterchamel2004, carter2005covariant, carter2005covIII}, based on the Milne-Cartan structure of the space-time. This approach allows for a more direct comparison with the general relativistic results (see, e.g., \cite{chachoua2006,chamel2008two}) than the Newtonian expressions from classical mechanics and hydrodynamics within the traditional `3+1' space-time decomposition, as developed, e.g., by \cite{prix2004variational}. As shown in Appendix~\ref{app:newt_lim}, the comparison between the Newtonian limit of Eq.~\eqref{FORCE_BALANCE} and results from~\cite{sourie20glitch} leads to the following identification:
 \begin{equation}
 \label{eq:identification}
    \mathcal{R}= m_{\n} n_{\n} \kappa_{\n} \, \xi \ \ \text{and} \ \ \mathcal{A} = N_{\p}\, ,
\end{equation}
from which we deduce that $\mathcal{R}$ corresponds to the microscopic drag coefficient usually introduced in the literature (see, e.g., \cite{andersson2006mutual}). 
Using Eq.~\eqref{eq:identification}, the mutual-friction force~\eqref{FORCE_BALANCE} finally reads 
\begin{equation}
    f^{\text{mf}}_{\mu} =  n_{\n}^{\nu} \varpi^{\n}_{\nu \mu} =   n_{\n}  w^{\n} \xi \! \perp_{\mu\nu}u_{\p}^{\nu} - N_{\p}\, n_{\p}^{\nu}\varpi^{\n}_{\nu \mu}\, , 
    \label{mutual_friction_force}
\end{equation}
which generalizes equation~(76) from~\cite{langlois1998differential} to the case in which each neutron vortex line is pinned to $N_{\p}$ proton fluxoids.
Introducing the space-time metric $g^{\mu\nu}$ with signature $(-,+,+,+)$, the projector to the surface orthogonal to the vortex worldsheet is explicitly given by~\citep{langlois1998differential}
\begin{equation}
\label{perp}
\perp^{\mu}_{\nu}\, =\dfrac{1}{\left(w^{\n}\right)^2}\, g^{\mu\sigma}g^{\rho\tau}  \varpi^{\n}_{\sigma\tau}\, \varpi^{\n}_{\nu\rho} = \dfrac{1}{\left(w^{\n}\right)^2}\,   \varpi^{\n \mu  \rho}\, \varpi^{\n}_{\nu\rho} \, ,
\end{equation}
where the scalar   
\begin{equation}
\label{eq:vorticity_ampl}
w^{\n} = \sqrt{g^{\rho\mu}g^{\sigma\nu}\varpi^{\n}_{\mu\nu}\varpi^{\n}_{\rho\sigma}/2}
\end{equation}
is related to the average surface density of vortex lines by  
\begin{equation}
\label{eq:vortex_surfdens}
    \mathcal{N}_{\n} = \dfrac{w^{\n}}{m_{\n}\kappa_{\n}}\, .
\end{equation}

Let us remark that the microscopic expression of the drag-to-lift ratio $\xi$ is likely to depend on various physical parameters, for which suitable relativistic definition must be used. For instance, the temperature in the covariant formulation must be understood as the scalar given by Eq.~\eqref{eq:temperature}.

\section{Mutual friction in a (quasi-)stationary and axisymmetric space-time}
\label{sec:mutual-friction}

\subsection{Space-time symmetries and fluid 4-velocities}

We now restrict our study to (quasi-)stationary and axisymmetric space-times, as would be relevant for the modelling of rotating NSs (see, e.g., \cite{paschalidis2017} for a recent review). The validity of these assumptions is discussed at the beginning of Section~3 of~\cite{langlois1998differential}. In what follows, we thus assume that there exist two Killing vectors: $k^{\mu}$ for stationarity and $h^{\mu}$ for axisymmetry\footnote{While $h^{\mu}$ is an exact Killing vector, $k^{\mu}$ is only an \textit{approximate} Killing vector because of the small non-circular motion considered in the following. This means that $k^{\mu}$ satisfies the following condition 
\begin{equation}
    \nabla_{\mu} k_{\nu} + \nabla_{\nu} k_{\mu} = \mathcal{O}\left(\dfrac{1}{L}\right)\, ,
    \label{eq:quasi_killing}
    \end{equation}
where $L$ is a length-scale very large with respect to the stellar radius (see Eq.~(49) of \cite{langlois1998differential}).
}. 
The axisymmetry (resp. quasi-stationarity) of the fluid flows translates into the exact (resp. approximate) vanishing of the Lie derivative along $h^{\mu}$ (resp. $k^{\mu}$) of any tensor field $q$ associated with matter, i.e.,
\begin{equation}
    \Lie_{\vec h}q = 0\ \ \text{and}\ \ \Lie_{\vec k}q \simeq  0\, ,
\end{equation}
where $\Lie_{\vec u} q$ denotes the Lie derivative of $q$ along the vector $u^\mu$.

Furthermore, the normal fluid is assumed to be rigidly rotating, so that its corresponding 4-velocity reads
\begin{equation}
\label{up}
u_{\p}^{\mu} = \gamma_{\p} \left(k^{\mu}+ \Omega_{\p}h^{\mu}\right),
\end{equation}
where $\gamma_{\p}$ is a Lorentz-type factor 
and $\Omega_{\p}$ is the (uniform) angular velocity of the normal fluid. Because charged particles are essentially locked to the pulsar's magnetosphere, $\Omega_{\p}$ coincides with the observed angular velocity of the pulsar. The 4-velocity of the neutron superfluid is taken as\footnote{Note that Eqs.~\eqref{up} and~\eqref{un} are not orthogonal decompositions.}
\begin{equation}
\label{un}
u_{\n}^{\mu} = \gamma_{\n} \left(k^{\mu}+ \Omega_{\n}h^{\mu}+ \tilde{v}_{\n}^{\mu}\right),
\end{equation}
where $\gamma_{\n}$ is the relevant Lorentz-type factor and $\Omega_{\n}$ is the (a priori non-uniform) angular velocity of the neutron superfluid.  The last term 
allows for the possibility of a non-circular motion (expected to be very small, see~\cite{langlois1998differential}). The factors $\gamma_{\p}$ and $\gamma_{\n}$ are fixed by the normalization conditions
\begin{equation}
     g_{\mu\nu}u_{\n}^{\mu}u_{\n}^\nu = -1 \ \ \text{and} \ \  g_{\mu\nu}u_{\p}^{\mu}u_{\p}^\nu = -1\, .
     \label{eq:normalisation}
\end{equation}

Further details on the decompositions \eqref{up} and \eqref{un} are given in Appendix~\ref{app:4vel_fluid}.

\subsection{Evolution equations}

Let us now express the mutual-friction force~\eqref{mutual_friction_force} in terms of $\Omega_{\p}$, $\Omega_{\n}$ and $\tilde{v}_{\n}^{\mu}$. We follow here the same approach as \cite{langlois1998differential}. Let us start by contracting equation~\eqref{mutual_friction_force} with $\varpi^{\n \mu \nu}$. This leads to 
\begin{equation}
\label{balanceI}
w^{\n} \! \perp^{\nu}_{\ \mu}\! \left(n_{\n}^{\mu} + N_{\p}\, n_{\p}^{\mu}\right) = n_{\n}\, \xi \, \varpi^{\n \nu\mu}\, u_{\p \mu}\, ,
\end{equation}
where we have made use of equation~\eqref{perp} and the orthogonality property $\perp^{\rho}_{\ \mu}\! \varpi^{\n\mu\nu}= \varpi^{\n \rho\nu}$. 
Introducing 
\begin{equation}
\alpha_{\n} =  h^{\nu}p_{\nu}^{\n}\, ,
\end{equation}
which can be interpreted as the angular momentum per neutron (see Section~\ref{sec:fluid_ang_mom}), a first dynamical equation is obtained by contracting equation~\eqref{balanceI} with $h_{\nu}$, i.e.,
\begin{equation}
\label{dyn1}
w^{\n} \, h_{\perp}^{\mu} \left( n_{\n \mu} + N_{\p}\,  n_{\p \mu} \right)=- n_{\n} \, \xi  \, u_{\p}^{\mu} \, \nabla_{\mu}\alpha_{\n}\, , 
\end{equation}
where $\nabla_{\mu} \alpha_{\n} = w^{\n}_{\mu\nu}h^{\nu}$, as deduced from $\mathcal{L}_{\vec h}p^{\n}_{\mu}=0$, and we have introduced the short-hand notation $h_{\perp}^{\mu} =\ \perp^{\mu}_{\ \nu}\! h^{\nu}$. A second  dynamical equation is derived by contracting equation~\eqref{balanceI} with $w^{\n}_{\nu \rho} h^{\rho}$, which yields
 \begin{equation}
\label{dyn2}
\left(n_{\n}^{\mu} + N_{\p}\, n_{\p}^{\mu}\right) \nabla_{\mu}\alpha_{\n} = n_{\n}\, w^{\n} \, \xi\,  h_{\perp}^{\mu} u_{\p \mu} \, .
\end{equation}

Making use of equations~\eqref{up} and~\eqref{un}, the dynamical equations~\eqref{dyn1} and \eqref{dyn2} lead to 
\begin{align}
    (1+X)\, k_{\mu} h_{\perp}^{\mu} + \dfrac{\tilde{\xi}}{w^{\n}}\dot{\alpha}_{\n} &= - h_{\perp}^{\mu}\tilde{v}_{\n\mu} - \left(\Omega_{\n} + X \Omega_{\p}\right)  h_{\perp}^2\, , \label{sys1} \\
    - w^{\n}\, \tilde{\xi}\, k_{\mu} h_{\perp}^{\mu} +(1+X)\, \dot{\alpha}_{\n} &= - \tilde{v}_{\n}^{\mu}\nabla_{\mu}\alpha_{\n} +  w^{\n}\, \tilde{\xi}\, \Omega_{\p} h_{\perp}^2\, ,\label{sys2}
\end{align}
where 
\begin{equation}
\label{eq:h_perp_square}
h_{\perp}^2 = h_{\perp}^{\ \mu} h_{\perp\, \mu} = h_{\perp}^{\ \mu} h_{\mu} \, , 
\end{equation}
and 
\begin{equation}
\dot{\alpha}_{\n} = k^{\mu}\nabla_{\mu} \alpha_{\n}\, ,
\end{equation}
which is small but non-zero since $k^{\mu}$ is not an exact Killing vector. Note that we have used here the fact that $\Lie_{\vec h} \alpha_{\n} =h^{\mu}\nabla_{\mu}\alpha_{\n} = 0$. The quantities $\tilde{\xi}$ and $X$ appearing in equations~\eqref{sys1} and \eqref{sys2} are defined as 
\begin{equation}
    \tilde{\xi} = \dfrac{\gamma_{\p}}{\gamma_{\n}}\xi\simeq \xi \, , 
\end{equation}
and 
\begin{equation}\label{eq.def.X}
X = \dfrac{\gamma_{\p}}{\gamma_{\n}}\dfrac{n_{\p}}{n_{\n}} N_{\p} \simeq \dfrac{n_{\p}}{n_{\n}} N_{\p} \, ,
\end{equation}
where $\gamma_{\p}\simeq \gamma_{\n}$ to a very good approximation, since the neutron superfluid and the normal fluid are expected to be very close to corotation at any time, as suggested by the very small glitch amplitudes.

The determinant of the system~\eqref{sys1}$-$\eqref{sys2} being non-zero, i.e.,  $\left(1+X\right)^2 + \tilde{\xi}^2>0$, this system of equations can be inverted, which leads to
\begin{align}
\label{vort_velI}
-h_{\perp}^{-2}k_{\mu}h_{\perp}^{\ \mu} &= \left(1-\mathcal{B}'\right)\left(\Omega_{\n} + \Omega_{\p}\dfrac{\mathcal{B}'}{1-\mathcal{B}'} + \Omega_{-}\right)\, ,  \\
\label{vort_velII}
\left(w^{\n}\right)^{-1}h_{\perp}^{-2}\dot{\alpha}_{\n} &= \mathcal{B}\left(\Omega_{\p} - \Omega_{\n} - \Omega_{+}
\right)\, ,
\end{align}
where we have introduced the following mutual-friction coefficients
\begin{equation}\label{eq:friction-coefs}
    \mathcal{B} = \dfrac{\tilde{\xi}}{\tilde{\xi}^2+\left(1+X\right)^2} \ \  \text{and} \ \    1-\mathcal{B}' = \dfrac{1+X}{\tilde{\xi}^2+\left(1+X\right)^2} \, ,
\end{equation}
in a similar manner to what has been done in the Newtonian framework (see Eq.~(57) of~\cite{sourie20force}) and  the (small) non-circular  contributions read 
\begin{align}
\Omega_{-} &= h_{\perp}^{-2}\tilde{v}_{\n}^{\mu} \left( h_{\perp\, \mu} -\tilde{\xi}  \left(w^{\n}\right)^{-1}\left(1+X\right)^{-1} \nabla_{\mu} \alpha_{\n} \right)\, ,  \label{omega_-}
\\
\Omega_{+} &= h_{\perp}^{-2}\tilde{v}_{\n}^{\mu} \left( h_{\perp\, \mu} + \tilde{\xi}^{-1}  \left(w^{\n}\right)^{-1}\left(1+X\right) \nabla_{\mu} \alpha_{\n} \right)\, .\label{omega_+}
\end{align}

The previous equations generalize those derived by \cite{langlois1998differential} to allow for  vortex pinning in the outer core of NSs. Indeed, taking $N_{\p} = 0$ (or equivalently $X=0$) in these equations leads to Eqs.~(84)$-$(87) of~\cite{langlois1998differential}, $\tilde{\xi}$ reducing in this case to the drag-to-lift ratio in the absence of pinning, denoted by $c_\text{r}$ in~\cite{langlois1998differential}.

As shown in Appendix~\ref{app:vel_vort}, the terms on the left-hand side of equations~\eqref{vort_velI} and~\eqref{vort_velII} can be respectively interpreted as the mean angular velocity of vortices (modulo a small non-circular contribution) and the (inwards) `radial' velocity of the vortex lines, i.e., along the unit vector $-\hat{r}^{\mu}$, where $\hat{r}^{\mu}$ is defined by
\begin{equation}
\label{eq:hat_r}
\hat{r}^{\mu} = \left(w^{\n}\right)^{-1} h_{\perp}^{-1} \varpi^{\n \mu\nu}h_{\nu}\ \  \text{and} \ \   \hat{r}^\mu \hat{r}_\mu = 1\, .
\end{equation} 
Note that $\hat{r}^{\mu}h_{\perp\, \mu} = 0$ and  $\perp^{\mu}_{\nu}\hat{r}^{\nu}=\hat{r}^{\mu}$, from which we deduce that the unit vectors $\hat{r}^{\mu}$ and  $\hat{h}_{\perp}^{\mu}=h_{\perp}^{\mu}/h_{\perp}$ form an orthonormal basis of the two-dimensional surface orthogonal to the vortex worldsheet. We thus have 
\begin{equation}
\perp^{\mu}_\nu\,= \hat{h}_{\perp}^{\mu}\hat{h}_{\perp \nu} + \hat{r}^{\mu}\hat{r}_{\nu} \, .
\end{equation}
The projector tangential to the vortex worldsheet is thus given by 
\begin{equation}
\eta^\mu_\nu \equiv \delta^\mu_\nu - \perp^{\mu}_\nu\,  = \delta^\mu_\nu - \hat{h}_{\perp}^{\mu}\hat{h}_{\perp \nu} - \hat{r}^{\mu}\hat{r}_{\nu} \, .
\end{equation}
The non-vanishing of the radial component of the mean vortex velocity (albeit very small) actually highlights the non-exact stationarity of the space-time considered here. 

Finally, the Newtonian limit of equations~\eqref{vort_velI}$-$\eqref{vort_velII} is found to match perfectly with results obtained from~\cite{sourie20force}, as shown in Appendix~\ref{app:newt_vel_vort}.

\subsection{Mutual-friction force in stationary rotating neutron stars}

Since $\eta^{\mu}_{\nu} f^{\text{mf}}_{\mu} =0 $, see Eqs.~\eqref{force_voln_bis} and \eqref{eq.force.worldsheet2}, the only non-zero components of the mutual-friction force~\eqref{mutual_friction_force} are those along the unit vectors $\hat{h}_{\perp}^{\mu}$ and $\hat{r}^{\mu}$. Substituting Eq.~\eqref{un} in \eqref{mutual_friction_force} recalling $\nabla_{\mu} \alpha_{\n} = w^{\n}_{\mu\nu}h^{\nu}$ and using Eqs.~\eqref{vort_velI} and \eqref{vort_velII}  leads to 
\begin{align}
  \hat{h}_{\perp}^{\mu} f^{\text{mf}}_{\mu} &= h^{\mu}f^{\text{mf}}_{\mu}/ h_{\perp}
\nonumber\\
  &=\gamma_{\n}n_{\n}\left(k^{\mu} + \Omega_{\n}h^{\mu} + \tilde{v}_{\n}^{\mu}\right)\nabla_{\mu}\alpha_{\n}/ h_{\perp}\nonumber\\
  &= \gamma_{\n}n_{\n}\left(\dot{\alpha}_{\n} + \tilde{v}_{\n}^{\mu}\nabla_{\mu}\alpha_{\n}\right) / h_{\perp}\nonumber\\
  &= \gamma_{\n} n_{\n} w^{\n} h_{\perp}\mathcal{B} \left[\Omega_{\p} - \Omega_{\n}+  X \Omega_{+}-\left(1+X\right) \Omega_{-} \right]\, , 
\label{mf_h}
\end{align}
and 
\begin{align}
  \hat{r}^{\mu} f^{\text{mf}}_{\mu} &= - \gamma_{\n} n_{\n}w^{\n} \left(k^{\mu} + \Omega_{\n}h^{\mu} + \tilde{v}_{\n}^{\mu}\right) h_{\perp\mu} / h_{\perp}\nonumber\\
  &= - \gamma_{\n} n_{\n}w^{\n} h_{\perp}\left(h_{\perp}^{-2}k_{\mu}h_{\perp}^{\mu} + \Omega_{\n} + h_{\perp}^{-2}\tilde{v}_{\n}^{\mu}h_{\perp\mu}\right) \nonumber\\
  &= \gamma_{\n} n_{\n} w^{\n} h_{\perp}\mathcal{B}'\left[\Omega_{\p} - \Omega_{\n} -  \dfrac{\tilde{\xi}^2\, \Omega_{+} +X\left(1+X\right) \Omega_{-}}{\tilde{\xi}^2+ X\left(1+X\right)}\right]\, . \label{mf_r}
\end{align}
Equations~\eqref{mf_h} and \eqref{mf_r} establish the expression for the mutual-friction force as a function of the fluid velocities only. In the limit $X=0$, equation~\eqref{mf_h} coincides with equation (89) from~\cite{langlois1998differential}. The Newtonian limit of Eqs.~\eqref{mf_h} and~\eqref{mf_r} are also found to be in perfect agreement with the expression~(5) from~\cite{sourie20glitch}, as shown in Appendix~\ref{app:newt_vel_vort}.

\subsection{Fluid angular momenta}
\label{sec:fluid_ang_mom}

In order to illustrate the impact of mutual-friction forces on the superfluid dynamics of NSs, let us now focus on the angular momentum transfer that takes place during the spin-up stage of pulsar glitches.

The existence of an (exact) Killing vector associated with axisymmetry allows for a gauge-invariant definition of the stellar angular momentum using the so-called Komar definition. 
In the usual `3+1' decomposition of the space-time, in which the space-time is foliated by a family $\left(\Sigma_t\right)_{t\in\mathds{R}}$ of spacelike hypersurfaces, the total angular momentum is thus defined as (see e.g. \cite{gourgoulhon2012})
\begin{equation}
\label{mom_komar}
J = - \int_{\Sigma_t} \left[ T_{\mu\nu}\eul^{\mu}h^{\nu} -\dfrac{1}{2}g^{\mu\nu}T_{\mu\nu}\eul_{\rho}h^{\rho}\right] \text{d}\Sigma\, ,
\end{equation}
where $T_{\mu\nu}$ is the energy-momentum tensor of the two-fluid system~\eqref{eq:Tmunu},  $\eul^{\mu}$ is the unit (future-oriented) vector normal to $\Sigma_{t}$ and $\text{d}\Sigma$ is the  volume element on the hypersurface $\Sigma_t$.  In a coordinate system adapted to the foliation (of the kind $x^0 = t, x^1, x^2, x^3$), we have $\text{d}\Sigma=\sqrt{\gamma}\, \text{d}x^1\text{d}x^2\text{d}x^3$, where $\gamma$ is the determinant of the spatial metric $\gamma_{ij}$ (using Latin letters for space indices $i,j=1,2,3$), defined from the 4-metric $\gamma_{\mu\nu}$ induced by $g_{\mu\nu}$ on $\Sigma_t$ (i.e., $\gamma_{ij} = g_{ij}$). The second term in Eq.~\eqref{mom_komar} can be dropped if the hypersurface is further required to be axisymmetric, i.e., if the Killing vector $h^{\mu}$ is tangent to $\Sigma_t$ such that $h^{\mu}\eul_{\mu}=0$.  
The angular momentum of the star thus simplifies as
\begin{equation}
J = \int_{\Sigma_t} j^{\mu}  \text{d}\Sigma_{\mu}\, ,
\end{equation}
where $\text{d}\Sigma_{\mu} = -\eul_{\mu}\text{d}\Sigma$ and $j^{\mu} = T^{\mu}_{\nu}h^{\nu} $ is the local angular momentum current. 
In presence of external forces ($f^{\text{ext}}_{\mu} \neq 0$), the time variation of the stellar angular momentum (as given from the time-translation generator $k^{\mu}$) is non-zero and reads
\begin{equation}
    \dfrac{\text{d}J}{\text{d}t} = \Gamma^{\text{ext}}\, , \ \ \ \ \ \Gamma^{\text{ext}} =   \int_{\Sigma_t} h^{\mu}f^{\text{ext}}_{ \mu}\,k^{\nu}\text{d}\Sigma_ {\nu}\, ,
\end{equation}
where $\Gamma^{\text{ext}} $ denotes the torque associated with the external force $f^{\text{ext}}_{ \mu}$~\citep{langlois1998differential}.

Making use of equation~\eqref{eq:Tmunu}, the local angular momentum current can be decomposed into
\begin{equation}
    j^{\mu} = j_{\n}^{\mu} + j_{\p}^{\mu} + \tilde{j}^{\mu}\, , 
\end{equation}
where we have introduced the following notations
\begin{equation}
   j_{\n}^{\mu} = n_{\n}^{\mu}p^{\n}_{\nu}h^{\nu} = \alpha_{\n} n_{\n}^{\mu}\, ,
       \label{ang_mome_curr_n}
\end{equation}
\begin{equation}
    j_{\p}^{\mu} = \left(n_{\p}^{\mu}p^{\p}_{\nu} + s\Theta u_{\p}^{\mu}u_{\p\nu}\right)h^{\nu}\, , 
\end{equation}
and 
\begin{equation}
    \tilde{j}^{\mu} = \Psi h ^{\mu}\, .
\end{equation}
Although the pressure term $ \tilde{j}^{\mu} $ does not allow for an unambiguous decomposition of the total angular momentum current in terms of separate fluid contributions, the total angular momentum of the star can still be unambiguously decomposed into a neutron and a normal parts since $ \tilde{j}^{\mu}\eul_{\mu} = 0$, yielding~\citep{langlois1998differential, sourie2016numerical}
\begin{equation}
    J = J_{\n} + J_{\p}\, , 
\end{equation}
where 
\begin{equation}
    J_{\n}=  \int_{\Sigma_t} j_{\n}^{\mu} \,   \text{d}\Sigma_{\mu}\, , \ \ \text{and} \ \   J_{\p}=  \int_{\Sigma_t} j_{\p}^{\mu} \,   \text{d}\Sigma_{\mu}\, .
\end{equation}

\subsection{Angular momentum transfer induced by mutual-friction forces}

Due to mutual-friction forces, angular momentum is redistributed between the initially decoupled neutron superfluid and the rest of the star during the rise of pulsar glitches. The corresponding dynamical equations read~\citep{langlois1998differential}
\begin{equation}\label{angular_momentum_transfer}
\dfrac{\text{d}J_{\n}}{\text{d}t} = \Gamma_{\text{mf}} \ \ \text{and} \ \ \dfrac{\text{d}J_{\p}}{\text{d}t} = \Gamma_{\text{ext}}-\Gamma_{\text{mf}}\, , 
\end{equation}
where
\begin{equation}
\label{gamma_int_RG}
\Gamma_{\text{mf}}  = \int_{\Sigma_t} h^{\mu}f_{\n \mu}\,k^{\nu}\text{d}\Sigma_ {\nu} = \int_{\Sigma_t} h^{\mu}f^{\text{mf}}_{ \mu}\,k^{\nu}\text{d}\Sigma_ {\nu}\, .
\end{equation}
Using Equation~\eqref{mf_h}, the mutual-friction torque 
is found to be given by 
\begin{align}\label{gamma_int_RG2}
    \Gamma_{\text{mf}}  &= \int_{\Sigma_t} \mathcal{B}\, \gamma_{\n} n_{\n} w^{\n} h_{\perp}^2 \left[\Omega_{\p} - \Omega_{\n}+  X \Omega_{+}-\left(1+X\right) \Omega_{-} \right]\,k^{\nu}\text{d}\Sigma_ {\nu}\ .
\end{align}
The pinning of proton fluxoids to neutron vortices does not only lead to a rescaling of the mutual-friction coefficient $\mathcal{B}$, but also introduces additional terms  in the mutual-friction torque through the parameter $X$, which in turn is proportional to the number $N_{\p}$ of  fluxoids attached to each vortex. These terms are still present in the Newtonian limit but were not taken into account in our previous analysis~\citep{sourie20glitch}, see also Appendix~\ref{app:newt_vel_vort}.

\section{Conclusions}

Following the seminal work of \cite{langlois1998differential}, we have adapted to the general-relativistic framework our recent Newtonian treatment~\citep{sourie20force} of the smooth-averaged mutual-friction force acting on the neutron superfluid and locally induced by the pinning of quantized neutron vortices to proton fluxoids in the outer core of superfluid NSs (or alternatively by the formation of vortex clusters of the kind  proposed by \cite{sedrakian1995superfluid}). Quite generally, the force can be written as $f^{\text{mf}}_{\mu} =  \mathcal{N}_{\n}\,   \mathcal{R}\,  u_{\p \perp \mu} + \mathcal{A} \, n_{\p}^{\nu}\varpi^{\n}_{\mu \nu}$. The apparent simplicity of this expression is deceptive 
because the microscopic coefficients $\mathcal{R}$ and $\mathcal{A}$ may depend on various physical parameters, 
such as the temperature~\eqref{eq:temperature}, the currents through the scalars $n_{\n}^\mu n_{\n \mu}$, $n_{\p}^\mu n_{\p \mu}$, $n_{\n}^\mu n_{\p \mu}$, or the vorticity~\eqref{eq:vorticity_ampl}. \cite{langlois1998differential} implicitly  assumed $\mathcal{A}=0$. However, this assumption turns out to be unrealistic whenever vortices interact with fluxoids.

Using the fully 4D covariant formulation of Newtonian dynamics of \cite{carterchamel2004,carter2005covariant,carter2005covIII}, we have shown how to relate the global general-relativistic dynamics of superfluid NSs to the local nonrelativistic dynamics of individual vortices. Comparing with our study of the nonrelativistic motion of a single vortex \citep{sourie20force}, we have thus been able to identify $\mathcal{R}$ with the drag coefficient and $\mathcal{A}$ with the mean number $N_{\p}$ of fluxoids pinned to each vortex.

According to our recent Newtonian study \citep{sourie20glitch}, vortex pinning may have important implications for the dynamics of superfluid NSs. 
Considering quasi-stationary and axisymmetric rotating NSs, we have derived the 
general-relativistic dynamical equations describing the mean motion of vortices, Eqs.~\eqref{vort_velI} and \eqref{vort_velII}, 
as well as the transfer of angular momentum between the different stellar 
components, Eqs.~\eqref{angular_momentum_transfer} and \eqref{gamma_int_RG2}. This set of equations generalizes that obtained earlier by \cite{langlois1998differential}. The role of the pinning of fluxoids to vortices or the formation of vortex clusters on the global dynamics of NSs, as embedded in the new parameter $X$ defined by Eq.~\eqref{eq.def.X}, appears to be rather nontrivial.

The present work provides all the necessary equations for carrying out realistic simulations of cold superfluid NSs in full general relativity  allowing for vortices and fluxoids, as required for the detailed interpretation of pulsar frequency glitches. 

\section*{Acknowledgements}

This work was supported by the Fonds de la Recherche Scientifique (Belgium) under grants no. 1.B.410.18F, CDR J.0115.18, and PDR T.004320. Partial support comes also from the European Cooperation in Science and Technology (COST) action CA16214. 

\section*{Data Availability}

No new data were generated or analysed in support of this research.

\appendix

\section{Newtonian limit of the force balance equation~\eqref{FORCE_BALANCE}}
\label{app:newt_lim}

In the present appendix, we use the covariant Newtonian formulation developed by Carter and Chamel~\citep{carterchamel2004, carter2005covariant, carter2005covIII} to match the global general-relativistic description with the local nonrelativistic dynamics of individual vortices discussed in \cite{sourie20force}. We will write the speed of light $c$ explicitly since we are interested in the limit $c\rightarrow+\infty$. 

\subsection{Covariant Newtonian formulation for superfluid neutron stars}

 Let $t_{\mu}\equiv \partial_\mu t$ be the gradient of the preferred Newtonian time coordinate $t$ associated with the foliation of the space-time into flat three-dimensional spaces with coordinates $X^i$ (we shall use Latin letters for spatial indices). Introducing the four-dimensional symmetric covariant tensor $\gamma_{\mu\nu}$, as obtained from pulling back the Euclidean spatial metric $\gamma_{ij}$, the (locally flat) Lorentzian metric of the relativistic description can be expressed as \citep{chachoua2006}
\begin{equation}\label{eq.metric.Newt.cov}
    g_{\mu\nu} = \gamma_{\mu\nu} - c^2 t_\mu t_\nu \, . 
\end{equation}
The factor $c^2$ is introduced here because in an ``Aristotelian'' coordinate system corresponding to the usual space-time decomposition, $x^0$ coincides with the Newtonian time $t$ (not $c t$) and $x^i$ with the space coordinates $X^i$. The tensor $\gamma_{\mu\nu}$ is not a space-time metric since it is degenerate
 \begin{equation}
 \label{eq:gammamunu_enu}
 \gamma_{\mu\nu}e^{\nu} = 0\, , 
 \end{equation}
 thus defining the so-called ether-frame flow vector $e^{\mu}$, normalized as
\begin{equation}
\label{eq:emu_tmu}
e^{\mu} t_{\mu} = 1\, .
\end{equation}
Similarly, the contravariant metric tensor can be expressed as \citep{chachoua2006}
\begin{equation}\label{eq.metric.Newt.cont}
    g^{\mu\nu} = \gamma^{\mu\nu} - \frac{1}{c^2}e^\mu e^\nu \, , 
\end{equation}
where the symmetric contravariant tensor $\gamma^{\mu\nu}$ is obtained from a pushforward of the three-dimensional Euclidean spatial metric $\gamma^{ij}$. 
Note that $\gamma_{\mu\nu}$ and $\gamma^{\mu\nu}$ satisfy the following relation: 
\begin{equation}
 \label{eq:gammamunu_gammamunu}
\gamma^{\mu\nu}\gamma_{\nu \rho} \equiv \gamma^{\mu}_{\rho} = \delta^{\mu}_{\rho} - e^{\mu}t_\rho\, , 
\end{equation} 
where $\delta^{\mu}_{\rho}$ denotes the Kronecker symbol. Like $\gamma_{\mu\nu}$, 
the tensor $\gamma^{\mu\nu}$ is also degenerate
 \begin{equation}
 \label{eq:gammamunu_tnu}
    \gamma^{\mu\nu}t_{\nu} = 0\, . 
\end{equation}
Therefore, the Newtonian space-time is characterized by the absence of any metric. 
This means that indices cannot be lowered or raised. In other words, covariant and contravariant indices of any space-time tensor are intrinsic. A special care is therefore needed when taking the Newtonian limit of relativistic expressions.

The fully antisymmetric four-dimensional 
contravariant spatial measure tensor $\varepsilon^{\mu\nu\rho}$ is obtained by pushforward of the Euclidean measure tensor $\varepsilon^{ijk}$. 
The covariant spatial measure tensor is given by 
\begin{equation}
\varepsilon_{\mu\nu\rho} = \gamma_{\mu\alpha}\, \gamma_{\nu\beta}\,\gamma_{\rho\gamma}\,\varepsilon^{\alpha\beta\gamma}\, .
\end{equation}
The fully covariant space-time measure tensor $\varepsilon_{\mu\nu\rho\sigma}$ is defined by the relation 
\begin{equation}
t_\mu=\frac{1}{3!}\varepsilon_{\mu\nu\rho\sigma} \varepsilon^{\nu\rho\sigma}\, .
\end{equation}
The contravariant counterpart $ \varepsilon^{\mu\nu\rho\sigma}$ is obtained by the normalization condition 
\begin{equation}
\varepsilon_{\mu\nu\rho\sigma} \varepsilon^{\mu\nu\rho\sigma}=-4!\, .
\end{equation}
Note that we have
\begin{equation}
\label{eq:3_eps_t_e}
    \varepsilon^{\mu\nu\rho}t_{\rho} = 0\, , \ \ \varepsilon_{\mu\nu\rho}e^{\rho} = 0\, , \text{and} \ \ \varepsilon^{\mu\nu\rho\sigma}t_{\sigma} = \varepsilon^{\mu\nu\rho} \, .
\end{equation}

\subsection{Force balance equation}

Making use of Eq.~\eqref{eq.metric.Newt.cov}, the relativistic dissipative force~\eqref{drag_force} can be written as
\begin{equation}
f^{\n}_{\text{d}\,\mu} =  \mathcal{C}_r \, (\gamma_{\mu\nu}-c^2 t_\mu t_\nu) \perp^\nu_\rho u^\rho_{\p} \, .
 \label{drag_force-Newt}
\end{equation}
Substituting Eq.~\eqref{perp} in Eq.~\eqref{drag_force-Newt} using Eq.~\eqref{eq.metric.Newt.cont} yields 
\begin{align}
    f^{\n}_{\text{d}\,\mu} &= \dfrac{\mathcal{C}_r}{\left(w^{\n}\right)^2}\,  \left(\gamma_{\mu\nu} - c^2t_{\mu}t_{\nu} \right) \left(\gamma^{\nu\lambda}\gamma^{\sigma\tau} \varpi^{\n}_{\tau\lambda}\varpi^{\n}_{\sigma\rho} \right.\nonumber\\
     &\ \ \ \ \ \left.-\gamma^{\nu\lambda} \dfrac{e^{\sigma}e^{\tau}}{c^2}\varpi^{\n}_{\tau\lambda}\varpi^{\n}_{\sigma\rho} - 
     \gamma^{\sigma\tau} \dfrac{e^{\nu}e^{\lambda}}{c^2}\varpi^{\n}_{\tau\lambda}\varpi^{\n}_{\sigma\rho}\right)u^\rho_{\p} \, .
\end{align}
In the Newtonian limit $c\rightarrow+\infty$, this becomes 
\begin{equation}
   f^{\n}_{\text{d}\,\mu}  = \dfrac{\mathcal{C}_r}{\left(w^{\n}\right)^2}\, \left(\gamma_{\mu}^{\lambda}\gamma^{\sigma\tau} \varpi^{\n}_{\tau\lambda}\varpi^{\n}_{\sigma\rho} + t_\mu e^{\lambda}  \gamma^{\sigma\tau} \varpi^{\n}_{\tau\lambda}\varpi^{\n}_{\sigma\rho}\right)  u^\rho_{\p} \, .
   \label{eq:newt_diss_force}
\end{equation}

Let $w_{\n}^{\mu}$ be the neutron vorticity vector, defined as \citep{carter2005covIII}
\begin{equation}
    w_{\n}^{\mu}= \dfrac{1}{2}\varepsilon^{\mu\nu\rho\sigma}\varpi^{\n}_{\nu\rho}t_{\sigma} = \dfrac{1}{2}\varepsilon^{\mu\nu\rho}\varpi^{\n}_{\nu\rho} \, .
    \label{eq:newt_vorticity_vec}
\end{equation}
It follows from Eq.~\eqref{eq:3_eps_t_e} that 
\begin{equation}
 w_{\n}^{\mu}t_\mu = 0\, , 
\end{equation}
meaning that the vorticity vector $w_{\n}^{\mu}$ is purely spatial. Its norm is
given by 
\begin{equation}
    \gamma_{\mu\nu} w_{\n}^{\mu} w_{\n}^{\nu} = \dfrac{1}{2}\gamma^{\mu\nu}\gamma^{\rho\sigma}\varpi^{\n}_{\mu\rho}\varpi^{\n}_{\nu\sigma} =  \left(w^{\n}\right)^2\, , 
    \end{equation}
where the vorticity scalar $w^{\n}$ is the Newtonian limit ($c\rightarrow +\infty$) of Eq.~(\ref{eq:vorticity_ampl}).

\begin{figure}
\centering
\includegraphics[width = 0.3\textwidth]{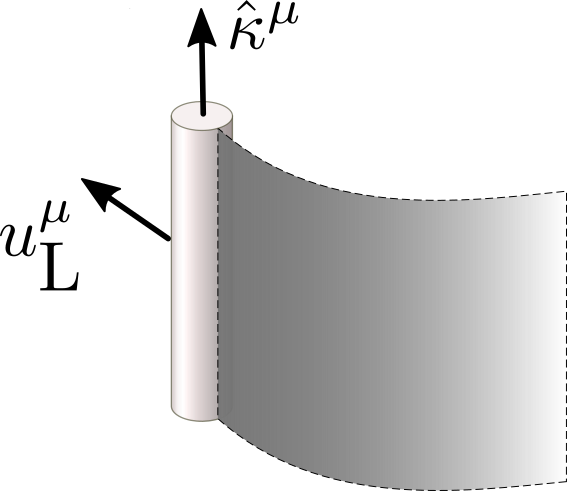}
\caption{The 4-vectors $u_{\vort}^{\mu}$ and $\hat{\kappa}^{\mu}$ form an orthogonal basis of the two-dimensional surface swept by the quantized vortex. See the text for details.}
\label{fig:vortex_newt}
\end{figure}

From the degeneracy of the vorticity 2-form $w^{\n}_{\mu\nu}$, we have 
\begin{equation}
    \varpi^{\n}_{\mu\nu} w_{\n}^{\nu}= 0 \, ,
\end{equation}
see, e.g., Eq.~(75) from~\cite{carter2005covIII} or Eq.~(41) from~\cite{chamel2006effect}. The unit spatial vector $\hat{\kappa}^{\mu}$, defined as
\begin{equation}
    \hat{\kappa}^{\mu} = \dfrac{w_{\n}^{\mu}}{w^{\n}}\, , \ \ \ \  \gamma_{\mu\nu}  \hat{\kappa}^{\mu}  \hat{\kappa}^{\nu} = 1 \, ,
    \label{eq:hat_kappa}
\end{equation}
can thus be seen as the unit vector along the vortex line. Since $w^{\n}_{\mu\nu}$ is of rank 2, we  can introduce a vector $u_{\vort}^{\mu}$ satisfying (see Eq.~(89) of \cite{carter2005covIII})
\begin{equation}
    \varpi^{\n}_{\mu\nu} u_{\vort}^{\nu} = 0\, , \ \ u_{\vort}^{\nu} t_{\nu} = 1\, , \ \ \text{and} \ \ \gamma_{\mu\nu}\,  u_{\vort}^{\mu}\hat{\kappa}^{\nu} = 0\, .
\label{eq:4vel_vort}
\end{equation}
The vorticity surface-generating 4-vector $u_{\vort}^{\mu}$ can be interpreted as the local average 4-velocity of the vortices (in the sense that the vorticity 2-form $w^{\n}_{\mu\nu}$ is Lie transported by $u_{\vort}^{\mu}$) in a direction orthogonal to $\hat{\kappa}^{\mu}$, as illustrated in Fig.~\eqref{fig:vortex_newt}. 
We now define the spatial part of the vortex 4-velocity as 
\begin{equation}
\label{eq:3vel_vort}
    v_{\vort}^{\mu} = u_{\vort}^{\mu} - e^{\mu}\, .
\end{equation}
Indeed, the normalization condition $u_{\vort}^{\mu}t_{\mu} = 1$, in combination with equation~\eqref{eq:emu_tmu}, leads to 
\begin{equation}
\label{eq:vvelmu_tmu}
    v_{\vort}^{\mu} t_{\mu} = 0\, .
\end{equation}
Besides, inserting equation~\eqref{eq:3vel_vort} into the last relation in~\eqref{eq:4vel_vort} yields
\begin{equation}
\label{eq:ortho_vL}
     \gamma_{\mu\nu}\,  v_{\vort}^{\mu}\hat{\kappa}^{\nu} = 0\, .
\end{equation}
From the previous considerations, the neutron vorticity 2-form $\varpi^{\n}_{\mu\nu}$ is found to be expressible as~\citep{carter2005covIII}
\begin{equation}
\label{eq:newt_vorticity}
    \varpi^{\n}_{\mu\nu} = \left(\varepsilon_{\mu\nu\rho} + 2v_{\vort}^{\sigma}t_{\left[\nu\right.}\varepsilon_{\left.\mu\right]\rho\sigma}\right) w^{\n} \hat{\kappa}^{\rho}\, .
\end{equation}

Defining the spatial part of the fluid 4-velocities by
\begin{equation}
  v_{\n}^{\mu} = u_{\n}^{\mu} - e^{\mu}\, , \ \ \ \   v_{\n}^{\mu} t_{\mu} = 0\, ,
  \label{vn_newt}
\end{equation}
and 
\begin{equation}
  v_{\p}^{\mu} = u_{\p}^{\mu} - e^{\mu}\, , \ \ \ \   v_{\p}^{\mu} t_{\mu} = 0\, ,
\label{vp_newt}
\end{equation}
the force balance equation~\eqref{FORCE_BALANCE} can  be finally recast as 
\begin{align}
\label{eq:force_balance_long}
    0 =&\ \ n_{\n}w^{\n}\varepsilon_{\mu\nu\rho}\left(v_{\n}^{\nu} - v_{\vort}^{\nu}\right)\hat{\kappa}^{\rho} +  n_{\n}w^{\n} t_{\mu} \varepsilon_{\nu\rho\sigma} v_{\n}^{\nu} v_{\vort}^{\rho}\hat{\kappa}^{\sigma} \nonumber\\
    &+ \mathcal{N}_{\n} \, \mathcal{R} \, \gamma_{\mu\nu}\left(v_{\p}^{\nu} - \hat{\kappa}^{\nu}\hat{\kappa}_{\rho}v_{\p}^{\rho} - v_{\vort}^{\nu}\right) \nonumber\\
    &+ \mathcal{N}_{\n} \, \mathcal{R} \, t_{\mu} \, \gamma_{\nu\rho}v_{\vort}^{\nu}\left(v_{\vort}^{\rho} -v_{\p}^{\rho}\right) \nonumber\\
    &+ \mathcal{A}\,  n_{\p}w^{\n}\varepsilon_{\mu\nu\rho}\left(v_{\p}^{\nu} -v_{\vort}^{\nu}\right)\hat{\kappa}^{\rho} \nonumber\\
    &+  \mathcal{A}\, n_{\p}w^{\n} t_{\mu} \varepsilon_{\nu\rho\sigma} v_{\p}^{\nu} v_{\vort}^{\rho}\hat{\kappa}^{\sigma}\, ,
\end{align}
where we have used Eqs.~\eqref{eq:gammamunu_enu}, \eqref{eq:emu_tmu},  \eqref{eq:gammamunu_gammamunu},~\eqref{eq:3_eps_t_e}, \eqref{eq:newt_diss_force} and \eqref{eq:newt_vorticity}. Contracting this last relation with the space projection tensor, defined as
\begin{equation}
P^{\mu}_\lambda = \delta^{\mu}_\lambda - e^{\mu}t_{\lambda} = \gamma^{\mu}_\lambda\, ,
\end{equation}
yields (only space indices $i$, $j$, $k$ appear here)
\begin{align}
   0 = &\  -  \mathcal{N}_{\n}\, \rho_{\n}\, \kappa_{\n} \, \varepsilon_{ijk}\hat{\kappa}^j \left(v_{\n}^{k} - v_{\vort}^{k}\right) -  \mathcal{N}_{\n}\, \mathcal{A}\, \rho_{\p}\, \kappa_{\p}\,  \varepsilon_{ijk}\hat{\kappa}^j \left(v_{\p}^{k} - v_{\vort}^{k}\right) \nonumber\\
   &+  \mathcal{N}_{\n}\, \mathcal{R}\,  \gamma_{ij}\left(v_{\p}^{j} - \hat{\kappa}^{j}\hat{\kappa}_{k}v_{\p}^{k} - v_{\vort}^{j}\right)\, , 
   \label{eq:newt_force_balance}
\end{align}
 where $\rho_{\n} = m_{\n}\, n_{\n}$, $\rho_{\p} = m_{\p} \, n_{\p}$, $\kappa_{\p} = h / (2 m_{\p})$ and we have made use of equation~\eqref{eq:vortex_surfdens}.  This last expression coincides with the sum\footnote{Let us remark that only velocities orthogonal to the direction $\hat{\kappa}^i$ of the vortex lines were considered in~\cite{sourie20force, sourie20glitch}, leading to $\hat{\kappa}_k v_{\p}^k =0$.} of equations~(1), (2) and (3) from~\cite{sourie20glitch} multiplied by the mean vortex surface density $\mathcal{N}_{\n}$,  as expected from the vanishing of the total force acting on individual vortex lines, provided that 
 \begin{equation}
 \label{identification}
    \mathcal{R} = \rho_{\n} \kappa_{\n}\, \xi \ \ \text{and} \ \ \mathcal{A} = N_{\p}\, , 
 \end{equation}
where $\xi$ is the (microscopic) `drag-to-lift' ratio and $N_{\p}$ is the number of proton fluxoids pinned to each vortex line. Note that the time component of the force balance equation~\eqref{eq:force_balance_long}, as obtained by contracting with $e^{\mu}$, yields the same equation as Eq.~\eqref{eq:newt_force_balance} (projected along $v_{\vort}^i$).

\section{Fluid 4-velocities}
\label{app:4vel_fluid}

Without any loss of generality, the fluid 4-velocities can be expressed as 
\begin{equation}
    u_{\n}^{\mu} = u_{\n}^{1}\, k^{\mu} + u_{\n}^{2} \,h^{\mu} + u_{\n}^{1}\,V_{\n}^{\mu}\ \ \text{and} \ \ u_{\p}^{\mu} = u_{\p}^{1}\, k^{\mu} + u_{\p}^{2} \,h^{\mu} + u_{\p}^{1}\, V_{\p}^{\mu}\, ,
\end{equation}
where each  vector $V_{_X}^{\mu}$ satisfies $V_{_X}^{\mu}k_{\mu} = V_{_X}^{\mu}h_{\mu} = 0$.

Assuming quasi-circular motion, one has $V_{_X}\ll 1$ (recalling we set $c=1$), where $V_{_X} = \sqrt{V_{_X \mu} V_{_X}^{\mu}}$. Therefore, the strain-rate tensors of the fluids satisfy
\begin{equation}
    \nabla_{\mu} V_{_X \nu} + \nabla_{\nu} V_{_X \mu} \ll \dfrac{1}{R}\, , 
\end{equation}
i.e., $\nabla_{\left(\mu\right.} V_{_X\left.\nu\right)} = \mathcal{O}\left(L^{-1}\right)$, where $L\gg R$ ($R$ being the radius of the star). Since the quasi-Killing vector $k^{\mu}$ is also subject to a  similar condition, see Eq.~\eqref{eq:quasi_killing}, one can now define $\tilde{k}^{\mu} = k^{\mu}  + V_{\p}^{\mu}$, which in turn satisfies
    \begin{equation}
   \nabla_{\mu} \tilde{k}_{\nu} + \nabla_{\nu} \tilde{k}_{\mu} = \mathcal{O}\left(\dfrac{1}{L}\right)\, , 
\end{equation}
and $\tilde{k}^{\mu}\tilde{k}_{\mu} \simeq  k^{\mu}k_\mu$ at lowest order in $V_{_X}$.
The 4-vector $\tilde{k}^{\mu}$ can  thus be  interpreted as another quasi-Killing vector associated with stationarity (this reflects the gauge freedom in the definition of $k^{\mu}$ as a `quasi'-Killing vector). The fluid 4-velocities can thus be rewritten as 
\begin{equation}
    u_{\p}^{\mu} = \gamma_{\p}\left( \tilde{k}^{\mu} +  \Omega_{\p} h^{\mu}\right) \ \ \text{and} \ \ u_{\n}^{\mu} =\gamma_{\n} \left(\tilde{k}^{\mu} +  \Omega_{\n} h^{\mu} + \tilde{v}_{\n}^{\mu}\right) \, ,
\end{equation}
where $\gamma_{\p} = u_{\p}^{1}$, $\Omega_{\p}= u_{\p}^{2} / u_{\p}^{1}$, $\gamma_{\n}= u_{\n}^{1}$, $\Omega_{\n} = u_{\n}^{2} / u_{\n}^{1}$, and $\tilde{v}_{\n}^{\mu} = V_{\n}^{\mu} - V_{\p}^{\mu}$. The expression considered for the normal 4-velocity can be seen as a gauge fixing condition for the quasi-Killing vector associated with stationarity (absence of non-circular motion for the normal fluid). Considering cylindrical coordinates $(t, r, \theta, z)$ adapted to the space-time symmetries, i.e.,
\begin{equation}
    \tilde{k}^{\mu} = \partial_t^{\mu}\ \ \text{and} \ \ h^{\mu} = \partial_\theta^{\mu}\, , 
\end{equation}
we have $\Omega_{\p} = u_{\p}^{\theta}/u_{\p}^t$ and $\Omega_{\n} = u_{\n}^{\theta}/u_{\n}^t$, from which we deduce that $\Omega_{\p}$ and $\Omega_{\n}$ are to be interpreted as the angular velocities of the normal fluid and neutron superfluid as seen by a static observer located at spatial infinity.

\section{Mean vortex velocity}
\label{app:vel_vort}

The definition of a mean 4-velocity of vortices is not devoid of ambiguities, referring either to some given observers (as e.g. in \cite{gavassino20}) or to the structure of the space-time under consideration (as in Appendix~\ref{app:newt_lim}, see Eq.~\eqref{eq:4vel_vort}). In either case, such a velocity should be so defined as to leave the vorticity 2-form $w^{\n}_{\mu\nu}$ invariant by Lie-transport. 

In what follows, the mean 4-velocity of the vortices is expressed in a similar form to that of the neutron superfluid~\eqref{un}, i.e.,
\begin{equation}
\label{uv}
u_{\text{L}}^{\mu} = \gamma_{\text{L}} \left(k^{\mu}+ \Omega_{\vort}h^{\mu}+ \tilde{v}_{\vort}^{\mu}\right)\, ,
\end{equation}
where $\tilde{v}_{\vort}^{\mu}$ is a small non-circular contribution satisfying $\tilde{v}_{\vort \mu} k^{\mu} = \tilde{v}_{\vort \mu} h^{\mu} =0$,  $\Omega_{\vort}$ is the mean (non-uniform) angular velocity of the vortices and $\gamma_{\vort}$ is some Lorentz-type factor obtained using the normalization condition $u_{\vort}^{\mu}u_{\vort \mu} = -1$.

Requiring\footnote{This condition can always be imposed since $\varpi^{\n}_{\mu\nu}$ is of rank 2.} $\varpi^{\n}_{\mu\nu}u_{\vort}^\nu  = 0$ (thus ensuring that the vorticity 2-form is Lie transported along the vortex 4-velocity) leads to
\begin{equation}
\label{eq_uv}
k^{\mu}\varpi^{\n}_{\mu \nu}+  \Omega_{\vort}h^{\mu}\varpi^{\n}_{\mu \nu} + \tilde{v}_{\vort}^{\mu}\varpi^{\n}_{\mu \nu} = 0\, .
\end{equation} 
Contracting this relation with  $w^{\n \nu \rho}h_{\rho}$  now yields 
\begin{equation}
-h_{\perp}^{-2}k_{\mu}h_{\perp}^{\ \mu}  = \Omega_{\vort} + \tilde{v}_{\vort}^{\mu} \, h_{\perp  \mu}h_{\perp}^{-2} \, , 
\label{vel_vort1}
\end{equation}
where we have made use of equations~\eqref{perp} and~\eqref{eq:h_perp_square}, from which we deduce that the term on the left-hand side of equation~\eqref{vort_velI} is to be interpreted as the mean  angular velocity of the vortices (plus a small non-circular contribution). On the other hand, contracting equation~\eqref{eq_uv} with $h^{\nu}$ yields
\begin{equation}
\dot{\alpha}_{\n} = - \tilde{v}_{\vort}^{\, \mu} w^{\n}_{\mu\nu}h^{\nu}.
\label{eq:alpha_dot_app}
\end{equation}
Introducing the unit space-like vector $\hat{r}^{\mu} = w_{\n}^{-1}h_{\perp}^{-1}\varpi^{\n \mu\nu}h_{\nu}$~\eqref{eq:hat_r}, which is both orthogonal to the vortex worldsheet and to the vector $h_{\perp}^{\mu}$, we get  
\begin{equation}
w_{\n}^{-1}h_{\perp}^{-2}\dot{\alpha}_{\n} = - h_{\perp}^{-1}\tilde{v}_{\vort}^{\, \mu}\hat{r}_{\mu}\, . \label{vel_vort2}
\end{equation}
The left-hand side of equation~\eqref{vort_velII} thus corresponds to the opposite of the vortex velocity along the unit vector $\hat{r}^{\mu}$ (i.e., to the `inwards radial' velocity of the vortices), as divided by $h_{\perp}$.

\section{Newtonian limits of the mean vortex velocity and the mutual-friction force}
\label{app:newt_vel_vort}

In this appendix, we will write the speed of light $c$ explicitly since we will take the Newtonian limit $c\rightarrow+\infty$. 

\subsection{Mean vortex velocity and equation of motion}

The concept of Killing vectors remains relevant in the Newtonian space-time~\citep{carter2005covariant,chamel2015}. 
In the present case, the (quasi-)Killing vector $k^{\mu} = \partial_t^{\mu}$ associated with stationarity coincides with the ether flow vector $e^{\mu}$. In cylindrical coordinates $\left(r, \theta, z\right)$ adapted to the space-time symmetries, we have 
\begin{equation}
    h^{\mu} = \partial_\theta^{\mu} = r e_{\theta}^{\mu}\, \ \ \text{and} \ \ \hat{\kappa}^{\mu} = e_{z}^{\mu}\, ,
    \label{eq:Killing_newt_adapt}
\end{equation}
where $\hat{\kappa}^{\mu}$ introduced in Eq.~\eqref{eq:hat_kappa} is the unit vector along the vortex direction and ($e_r^{\mu}$, $e_\theta^{\mu}$, $e_z^{\mu}$) is the right-handed orthonormal spatial vector basis associated to the cylindrical coordinates. Note that $h^{\mu}t_{\mu} = 0 = \hat{\kappa}^{\mu}t_\mu$.

Using Eq.~\eqref{uv}, the mean 4-velocity of the vortices reads 
\begin{equation}
    u_{\vort}^\mu=\gamma_{\vort} \left(e^{\mu}+ \Omega_{\vort} h^\mu+\tilde{v}_{\vort}^{\mu}\right)\, . 
\end{equation}
The condition $\tilde{v}_{\vort}^{\mu}k_{\mu}=0$ translates into $ t_{\mu} \tilde{v}_{\vort}^{\mu}=0$, as demonstrated below: 
\begin{equation}
    0 = g_{\mu\nu}\tilde{v}_{\vort}^{\mu}k^{\nu}  = (\gamma_{\mu\nu} - c^2 t_{\mu}t_{\nu} ) \tilde{v}_{\vort}^{\mu}e^{\nu} = - c^2 t_{\mu} \tilde{v}_{\vort}^{\mu}\, ,
\end{equation}
where we have made use of Eqs.~\eqref{eq.metric.Newt.cov} and \eqref{eq:gammamunu_enu}. The normalization condition $u_{\vort}^{\mu}t_{\mu} = 1$ together with Eq.~\eqref{eq:emu_tmu} imply that $\gamma_{\vort}=1$. Since $\tilde{v}_{\vort}^{\mu}$ is purely spatial, the orthogonality condition  $\tilde{v}_{\vort \mu}h^\mu=0$ becomes  $\gamma_{\mu\nu}\tilde{v}_{\vort}^{\mu}h^\nu=0$. By definition, 
$\tilde{v}_{\vort}^{\mu}$ is also orthogonal to $\hat{\kappa}^\mu$, see Eq.~\eqref{eq:4vel_vort}, therefore $\tilde{v}_{\vort}^{i} = \tilde{v}_{\vort}^r e_r^i$ (using Latin letters for spatial indices). Finally, the spatial part~\eqref{eq:3vel_vort} of the mean 4-velocity of the vortices reads 
\begin{equation}
    v_{\vort}^i=r\Omega_{\vort}e_\theta^{i}+  \tilde{v}_{\vort}^r e_r^i\, .
    \label{eq:v_vort_newt}
\end{equation}

We shall now derive the Newtonian limit of Eqs.~\eqref{vort_velI} and~\eqref{vort_velII} describing the average motion of vortices. Making use of Eqs.~\eqref{perp}, \eqref{eq.metric.Newt.cov}, \eqref{eq:gammamunu_enu} and \eqref{eq.metric.Newt.cont}, 
we have 
\begin{align}
    h_\perp^2 &=g_{\mu\nu} h_\perp^\mu h^\nu \nonumber \\ 
    &=g_{\mu\nu} \perp^\mu_\rho h^\rho h^\nu \nonumber \\ 
    &=g_{\mu\nu} \dfrac{1}{\left(w^{\n}\right)^2}\, g^{\mu\lambda} g^{\sigma\tau} \varpi^{\n}_{\tau\lambda}\, \varpi^{\n}_{\sigma\rho} h^\nu h^\rho  \nonumber \\ 
    &=\gamma_{\mu\nu} \dfrac{1}{\left(w^{\n}\right)^2}\, \left(\gamma^{\mu\lambda}-\dfrac{e^\mu e^\lambda}{c^2}\right)\left( \gamma^{\sigma\tau}-\dfrac{e^\sigma e^\tau}{c^2}\right) \varpi^{\n}_{\tau\lambda}\, \varpi^{\n}_{\sigma\rho} h^\nu h^\rho  \nonumber \\ 
    &=\gamma_{\mu\nu} \dfrac{1}{\left(w^{\n}\right)^2}\, \gamma^{\mu\lambda}\left( \gamma^{\sigma\tau}-\dfrac{e^\sigma e^\tau}{c^2}\right) \varpi^{\n}_{\tau\lambda}\, \varpi^{\n}_{\sigma\rho} h^\nu h^\rho  \nonumber \\ 
    &= \dfrac{1}{\left(w^{\n}\right)^2}\, \left( \gamma^{\sigma\tau}-\dfrac{e^\sigma e^\tau}{c^2}\right) \varpi^{\n}_{\tau\lambda}h^\lambda\, \varpi^{\n}_{\sigma\rho} h^\rho \, .
\end{align}
Taking the Newtonian limit $c\rightarrow +\infty$ using Eq.~\eqref{eq:newt_vorticity} yields 
\begin{align}
    h_\perp^2  
    &= \dfrac{1}{\left(w^{\n}\right)^2}\, \gamma^{\sigma\tau} \varpi^{\n}_{\tau\lambda}h^\lambda\, \varpi^{\n}_{\sigma\rho} h^\rho  \nonumber \\
    &= \gamma^{\sigma\tau} \varepsilon_{\tau\lambda\delta}\hat{\kappa}^\delta \, 
    \varepsilon_{\sigma\rho\mu}\hat{\kappa}^\mu
    h^\lambda h^\rho  \nonumber \\ 
    &=\hat{\kappa}^\delta\hat{\kappa}^\mu
    h^\lambda h^\rho {\varepsilon^{\sigma}}_{\lambda\delta} \, 
    \varepsilon_{\sigma\rho\mu}\nonumber \\ 
         &=\hat{\kappa}^\delta\hat{\kappa}^\mu
    h^\lambda h^\rho \left(\gamma_{\lambda\rho}\gamma_{\delta\mu} -\gamma_{\lambda\mu}\gamma_{\delta\rho} \right)\nonumber \\ 
       &= \gamma_{\mu\nu}h^{\mu} h^{\nu} - \left(\gamma_{\mu\nu}h^{\mu}\hat{\kappa}^{\nu}\right)^2\, .
\end{align}
It follows from Eq.~\eqref{eq:Killing_newt_adapt} that 
\begin{equation}\label{eq.hperp.newt}
       h_\perp^2 =  \gamma_{\mu\nu}h^{\mu} h^{\nu} = r^2\, .
\end{equation}
Similarly, 
\begin{align}\label{eq.kdothperp.newt}
   k_{\mu} h_{\perp}^{\mu} &= g_{\mu\nu} k^{\nu} h_{\perp}^{\mu} \nonumber \\
   &= g_{\mu\nu} e^{\nu} \perp^{\mu}_{\rho} h^{\rho} \nonumber \\
   &= (\gamma_{\mu\nu} - c^2 t_{\mu}t_{\nu } )\,  e^{\nu}\! \perp^{\mu}_{\rho} h^{\rho} \nonumber \\
      &=  - c^2 t_{\mu}  \perp^{\mu}_{\rho} h^{\rho} \nonumber \\
 &=   \dfrac{1}{(w^{\n})^2} e^{\nu} \gamma^{\sigma\mu} \varpi^{\n}_{\nu\mu} \varpi^{\n}_{\rho\sigma} h^{\rho} \nonumber \\
  &= \gamma^{\sigma\mu} \varepsilon_{\mu\nu\lambda}v_{\vort}^{\nu} \hat{\kappa}^{\lambda} \varepsilon_{\rho\sigma\delta}  \hat{\kappa}^{\delta}  h^{\rho} \nonumber \\
    &=  {\varepsilon^{\sigma}}_{\nu\lambda} \varepsilon_{\rho\sigma\delta} v_{\vort}^{\nu} \hat{\kappa}^{\lambda}   \hat{\kappa}^{\delta}  h^{\rho} \nonumber \\
    &=-  {\varepsilon^{\sigma}}_{\nu\lambda} \varepsilon_{\sigma\rho\delta} v_{\vort}^{\nu} \hat{\kappa}^{\lambda}   \hat{\kappa}^{\delta}  h^{\rho} \nonumber \\
        &= - \left(\gamma_{\rho\nu}\gamma_{\lambda\delta}- \gamma_{\nu\delta}\gamma_{\rho\lambda}\right) v_{\vort}^{\nu} \hat{\kappa}^{\lambda}   \hat{\kappa}^{\delta}  h^{\rho} \nonumber \\
                &=  - \gamma_{\rho\nu}v_{\vort}^{\nu}   h^{\rho} \, .
\end{align}
Collecting Eqs.~\eqref{eq.hperp.newt} and \eqref{eq.kdothperp.newt}, the left-hand side of Eq.~\eqref{vort_velI} thus reduces in the Newtonian limit to 
\begin{equation}
-h_{\perp}^{-2}k_{\mu}h_{\perp}^{\ \mu}  =  \dfrac{v_{\vort}^{\theta}}{r}=\Omega_{\vort}\, .
\end{equation}
Using Eqs.~\eqref{eq:3_eps_t_e}, \eqref{eq:newt_vorticity}, \eqref{eq:alpha_dot_app} and \eqref{eq:Killing_newt_adapt}, we find 
\begin{align}
    \dot{\alpha}_{\n} &= - \tilde{v}_{\vort}^{\, \mu} w^{\n}_{\mu\nu}h^{\nu} \nonumber\\
    &= - \tilde{v}_{\vort}^{\, \mu}  w^{\n} \varepsilon_{\mu\nu\rho}h^{\nu} \hat{\kappa}^{\rho} \nonumber\\
        &= - \tilde{v}_{\vort}^{\, r}  w^{\n} r \, .
\end{align}
Therefore, the left-hand side of Eq.~\eqref{vort_velII} is given by
\begin{equation}
    w_{\n}^{-1}h_{\perp}^{-2}\dot{\alpha}_{\n} = - \dfrac{\tilde{v}_{\vort}^{\, r}}{r} \, , 
\end{equation}
where we have made use of Eq.~\eqref{eq.hperp.newt}. 

To find the Newtonian limit of Eqs.~\eqref{omega_-} and \eqref{omega_+}, we need to evaluate the scalars $\tilde{v}_{\n}^{\mu} \, h_{\perp  \mu}$ and  $\tilde{v}_{\n}^{\mu} \nabla_{\mu} \alpha_{\n}$.
Using Eq.~\eqref{eq.metric.Newt.cov} and the fact that $\tilde{v}_{\n}^{\mu}t_\mu=0$,  we have
\begin{align}
    \tilde{v}_{\n}^{\mu} \, h_{\perp  \mu} &=    \tilde{v}_{\n}^{\mu} \, g_{\mu\nu}\! \perp^{\nu}_{\rho}h^{\rho} \nonumber \\
    &=    \tilde{v}_{\n}^{\mu} \, (\gamma_{\mu\nu} - c^2 t_\mu t_{\nu} )\perp^{\nu}_{\rho}h^{\rho} \nonumber \\
    &= \tilde{v}_{\n}^{\mu} \, \gamma_{\mu\nu} \! \perp^{\nu}_{\rho}h^{\rho} \, .
\end{align}   
Making use of Eqs.~\eqref{perp} and \eqref{eq:newt_vorticity}, and taking the limit $c\rightarrow+\infty$, yield
  \begin{align}    
\tilde{v}_{\n}^{\mu} \, h_{\perp  \mu} 
      &=    \tilde{v}_{\n}^{\mu} \, \gamma_{\mu\nu} \dfrac{1}{(w^{\n})^2}   \gamma^{\nu\lambda} \gamma^{\sigma\delta} \varpi^{\n}_{\delta\lambda} \varpi^{\n}_{\sigma\rho} h^{\rho} \nonumber \\
&=    \tilde{v}_{\n}^{\mu} \,  \gamma^{\lambda}_\mu  \gamma^{\sigma\delta}  
            \varepsilon_{\delta\lambda\nu}\hat{\kappa}^{\nu} \varepsilon_{\sigma \rho\gamma}\hat{\kappa}^{\gamma} h^{\rho} \nonumber \\
&=    \tilde{v}_{\n}^{\lambda} \,   \gamma^{\sigma\delta}  
            \varepsilon_{\delta\lambda\nu}\hat{\kappa}^{\nu} \varepsilon_{\sigma \rho\gamma}\hat{\kappa}^{\gamma} h^{\rho} \nonumber \\
        &=\tilde{v}_{\n}^{\lambda} \,   
            {\varepsilon^{\sigma}}_{\lambda\nu} \varepsilon_{\sigma \rho \gamma} \hat{\kappa}^{\gamma}  \hat{\kappa}^{\nu} h^{\rho} \nonumber\\ 
                    &=    \tilde{v}_{\n}^{\lambda} \,  \left(\gamma_{\rho\lambda}\gamma_{\nu\gamma} - \gamma_{\lambda\gamma}\gamma_{\rho\nu}\right)\hat{\kappa}^{\gamma}  \hat{\kappa}^{\nu} h^{\rho} \nonumber \\
                    &=0 \, ,
    \end{align}
where the last equality follows from $\gamma_{\mu\nu} \tilde{v}_{\n}^\mu h^{\nu} =0$ and $\gamma_{\mu\nu}\hat{\kappa}^{\mu}h^{\nu} = 0$. 
Similarly, using Eqs.~\eqref{eq:3_eps_t_e}, \eqref{eq:newt_vorticity}, and \eqref{eq:Killing_newt_adapt}, we find 
\begin{equation}
        \tilde{v}_{\n}^{\mu} \nabla_{\mu} \alpha_{\n}  = \tilde{v}_{\n}^{\mu} w^{\n}_{\mu\nu}h^{\nu} =  \tilde{v}_{\n}^{\, r}  w^{\n} r\, .
\end{equation}

Equations~\eqref{omega_-} and~\eqref{omega_+} thus become in the Newtonian limit 
\begin{align}
\Omega_{-} &= - \dfrac{\tilde{\xi}}{1+X} \dfrac{\tilde{v}_{\n}^r}{r}\, , 
\\
\Omega_{+} &= \dfrac{1+X}{\tilde{\xi}}\dfrac{\tilde{v}_{\n}^r}{r}\, .
\end{align}
 
The Newtonian expressions for the dynamical equations~\eqref{vort_velI} and~\eqref{vort_velII} finally  read
\begin{align}
\dfrac{v_{\vort}^\theta}{r} &=\Omega_{\vort} =  \left(1-\mathcal{B}'\right)\left(\Omega_{\n} + \Omega_{\p}\dfrac{\mathcal{B}'}{1-\mathcal{B}'} - \dfrac{\tilde{\xi}}{1+X} \dfrac{\tilde{v}_{\n}^r}{r}\right)\, ,  \\
-\dfrac{v_{\vort}^r}{r} &=-\dfrac{\tilde{v}_{\vort}^r}{r} = \mathcal{B}\left(\Omega_{\p} - \Omega_{\n}   - \dfrac{1+X}{\tilde{\xi}}\dfrac{\tilde{v}_{\n}^r}{r}
\right)\, ,
\end{align}
in perfect agreement with the expressions obtained from equation~(56) of~\cite{sourie20force} using
\begin{equation}
\label{eq:circ_newt}
    v_{\n}^i =r\Omega_{\n}e_\theta^i + \tilde{v}_{\n}^{r}e_{r}^i\, , \ \ \text{and} \ \ v_{\p}^i =r\Omega_{\p}e_\theta^{i}\, .
\end{equation}

\subsection{Mutual-friction force and torque}

Let us first show that the unit vector $\hat{r}^{\mu}$ reduces to $e_r^{\mu}$ in the Newtonian limit. Starting from the definition~\eqref{eq:hat_r} and substituting Eq.~\eqref{eq.metric.Newt.cont}, we find   
\begin{align}
    \hat{r}^{\mu} 
    &= w_{\n}^{-1}h_{\perp}^{-1} g^{\mu\rho}g^{\nu\sigma}\varpi^{\n}_{\rho\sigma} h_{\nu} \nonumber\\
        &= w_{\n}^{-1}h_{\perp}^{-1} \left(\gamma^{\mu\rho} - \dfrac{e^{\mu}e^{\rho}}{c^2}\right)
   \left(\gamma^{\nu\sigma} - \dfrac{e^{\nu}e^{\sigma}}{c^2}\right)\varpi^{\n}_{\rho\sigma} h_{\nu}\, .
\end{align} 
Taking the limit $c\rightarrow +\infty$ leads to  
\begin{align}
   \hat{r}^{\mu} &= w_{\n}^{-1}h_{\perp}^{-1} \gamma^{\mu\rho}  \varpi^{\n}_{\rho\sigma} h^{\sigma} \nonumber\\
  &=w_{\n}^{-1}h_{\perp}^{-1} \gamma^{\mu\rho}  \varepsilon_{\rho\sigma\delta} w^{\n} \hat{\kappa}^{\delta} h^{\sigma} \nonumber\\ 
  &= e_r^{\mu} \, .
\end{align}
We have made use of  Eq.~\eqref{eq:Killing_newt_adapt} for the last equality. 
Using Eq.~\eqref{eq:vortex_surfdens}, the Newtonian limit of Eqs.~\eqref{mf_h} and~\eqref{mf_r} thus reduces to 
\begin{align}
  f^{\text{mf} \,\theta} &=\rho_{\n}\, \kappa_{\n} \, \mathcal{N}_{\n}\, \left[ \mathcal{B} r \left(\Omega_{\p} - \Omega_{\n}\right)+ \mathcal{B}'\tilde{v}_{\n}^r \right]\, ,
  \label{mf_r_newt}\\ 
 f^{\text{mf}\, r} &= \rho_{\n}\, \kappa_{\n}\, \mathcal{N}_{\n} \, \left[ \mathcal{B}' r\left(\Omega_{\p} - \Omega_{\n}\right) - \mathcal{B} \tilde{v}_{\n}^r \right]\, .
 \label{mf_theta_newt}
\end{align}
Equations~\eqref{mf_r_newt} and~\eqref{mf_theta_newt} agree perfectly with the radial and orthoradial components of the mutual-friction force obtained from Eq.~(5) of~\cite{sourie20glitch} with the prescription~\eqref{eq:circ_newt}. Using Eq.~\eqref{gamma_int_RG2}, the mutual-friction torque is given by the integral over the  volume of the star
\begin{align}\label{gamma_int_Newt}
    \Gamma_{\text{mf}}  = \int \, n_{\n} w^{\n} r^2 \left[\mathcal{B}(\Omega_{\p} - \Omega_{\n})+   \mathcal{B}'\dfrac{\tilde{v}_{\n}^r}{r} \right]\,\text{d}V\, .
\end{align}
recalling that $ k^{\mu} = e^{\mu}$ and $e^{\mu}t_{\mu} = 1$, and using the fact that $\text{d}\Sigma_{\mu} = t_{\mu} \text{d}V$. 

Assuming circular motion ($\tilde{v}_{\n}^r=0$), neglecting entrainment effects between the fluids ($w^{\n}=2\, m_{\n}\,  \Omega_{\n}$), and considering uniform mutual-friction coefficients and fluid angular velocities, Eq.~\eqref{gamma_int_Newt} reduces to ($I_{\n}$ is the moment of inertia of  the neutron superfluid)
\begin{align}\label{gamma_int_Newt2}
    \Gamma_{\text{mf}}  = 2\,  \mathcal{B} \, I_{\n} \, \Omega_{\n} \,  (\Omega_{\p} - \Omega_{\n})\  ,
\end{align}
in perfect agreement with Eqs.~(10) and (11) of \cite{sourie20glitch}.

\bibliographystyle{mnras}
\bibliography{biblio}

\bsp	
\label{lastpage}
\end{document}